\documentstyle[twocolumn,prb,aps]{revtex}
\begin{document}
\twocolumn[\hsize\textwidth\columnwidth\hsize\csname @twocolumnfalse\endcsname
\draft

\title{
Hall Drag in Correlated Double Layer Quantum Hall Systems}

\author{Kun Yang}

\address{
Condensed Matter Physics 114-36, California Institute of Technology,
Pasadena, California 91125}
\date{\today}
\maketitle

\begin{abstract}
We show that in the limit of zero temperature, double layer quantum Hall
systems exhibit a novel phenomena called Hall drag, namely a current 
driven in one layer induces a voltage drop in the other layer, 
in the direction perpendicular 
to the driving current. The two-by-two
Hall resistivity tensor is quantized and 
proportional to the ${\bf K}$ matrix that describes the topological order
of the quantum Hall state, even when the 
${\bf K}$ matrix contains a zero
eigenvalue, in which case the Hall conductivity tensor does not exist.
Relation between the present work and previous ones is also discussed.
\end{abstract}
]

The experimental discovery and theoretical understanding of the
fractional quantum Hall effect (FQHE)\cite{pg,dassarma}
is one of the most important progress
in condensed matter physics in the past fifteen years. Recently, much attention
has been focusing on FQHE in multicomponent systems\cite{gm}. Such 
components may be the spins of electrons which are not frozen out when the 
external magnetic field is not too strong, or layer indices in multi-layered
system. Novel physics such as even-denominator FQHE
states\cite{suen,eisenstein}, spin-ferromagnetism\cite{sondhi}, spontaneous
inter-layer coherence\cite{wenzee,ezawa,murphy,yang}, and
canted antiferromagnetism\cite{zheng} has been discovered in these
systems.

In multi-layered systems, the interactions and correlations of electrons
in different layers is crucial to the FQHE. Such correlations are
captured by Halperin's multi-component trial 
wave functions\cite{halperin}. However they are not easy to directly 
detect in usual transport measurements, which is the most heavily used 
method in experimental studies of FQHE. 

In a drag measurement\cite{gramila}, separate electric contacts are made to 
the electron gas in the two different layers, and electric current is forced
to flow in one layer (called the driving layer). This current will induce
a measurable voltage drop in the other layer (called the drag layer),
even though no current is flowing in it. In the absence of a magnetic field,
the drag voltage is in the opposite direction of the driving current,
and 
the transresistance, defined as the ratio between the drag voltage and the
driving current, reflects the density fluctuations of {\em individual}
layers, and
vanishes at $T=0$ if the coupling between the layers is weak enough\cite{zm}.
In the presence of a magnetic field perpendicular to the layers, 
a perpendicular component of the drag voltage, called Hall drag, is
in principle possible\cite{hu}. In this paper we will show that this is
indeed the case in correlated double layer FQHE systems, and the
trans-Hall-resistivity $\rho^{xy}_{\uparrow\downarrow}$ is {\em finite}
and {\em quantized} at
$T=0$ even though the normal longitudinal drag voltage vanishes; the
finite $\rho^{xy}_{\uparrow\downarrow}$ reflects the {\em interlayer 
electron-electron correlations} in the 
{\em ground state}. Our results may be expressed in the compact form of a
2 by 2 Hall resistivity tensor:
\begin{equation}
\rho^{xy}_{ij}={\bf K}_{ij}h/e^2,
\label{result}
\end{equation}
where $i$ and $j$ are layer indices, and ${\bf K}$ is a 2 by 2 matrix that
describes the topological order of the quantum Hall state.\cite{wen1}
As we will see below, the above result is valid even when the ${\bf K}$ matrix
contains a zero eigen value. There is no longitudinal voltage drop at zero 
temperature.

Using a Chern-Simons-Ginsburg-Landau (CSGL)
type of effective theory generalized to
double layer systems, Renn\cite{renn} argued that the Hall conductivity
tensor is ${e^2\over h}{\bf K}^{-1}$. Our results are consistent with his.
However in our work we show that our results may be derived {\em exactly} 
using known {\it microscopic} 
wave functions for special types of electron-electron
interaction. Also his approach formally breaks down when ${\bf K}$ contains
a zero eigen value, as the inverse of ${\bf K}$ does not exist in this case.
In our approach however, since we calculate the {\em resistivity} tensor
directly (in the microscopic calculation), we still obtain well defined 
answers. When ${\bf K}$ contains a zero eigen value, the system supports a
charge neutral gapless mode. Duan\cite{duan} suggested that such a model
gives rise to a Hall drag resistivity that is not quantized. It is clear
from our exact calculation below that this is {\em not} the case; the
Hall drag resistivity is quantized even when ${\bf K}$ contains a 
zero eigen value.

In the rest of the paper we will start by considering a special case where
the Halperin wave functions are the exact ground states of the system, 
in which the trans-Hall-resistivity may be calculated exactly using the
exact microscopic wave functions.
We then formally derive the expressions for
$\rho^{xy}_{\uparrow\downarrow}$ for edge currents under more general
conditions, using
the chiral Luttinger liquid theory.
We conclude with comments on the experimental implications of our results.
Throughout the paper we assume no tunneling is allowed between the layers.

We begin by considering the limit that the Landau level spacing 
$\hbar\omega_c\rightarrow\infty$, so that all electrons are in the 
lowest Landau level (LLL) and has zero kinetic energy (measured from 
${1\over 2} \hbar\omega_c$). 
In this limit, the electron-electron interaction may be parametrized 
by Haldane's pseudopotentials\cite{haldane} $U_l$, which are the interaction
energies of a pair of electron in a state with relative angular momentum $l$. 
We assume that the
intralayer pseudopotentials $U_l^{\uparrow\uparrow}
=U_l^{\downarrow\downarrow} > 0$ for $l < m$, and interlayer 
pseudopotentials $U_l^{\uparrow\downarrow} > 0$ 
for $l < n$; all other $U$'s are zero.
We also introduce a circularly symmetric
confining potential $V(r)$, which is zero in the bulk and
increases smoothly with the distance from
the origin $r$ near the edge of the disc. 
We assume the chemical potential of the electron gas
$\mu=\mu_\uparrow=\mu_\downarrow$ is much smaller than the nonzero $U$'s, 
so that the electron gas stays in the small $V$ region. In this case the
ground state of the system is exactly the Halperin $(mmn)$ wave function:
\begin{equation}
\Psi_{mmn}=\prod_{i<j}(z_i^\uparrow-z_j^\uparrow)^m
(z_i^\downarrow-z_j^\downarrow)^m\prod_{i,j}(z_i^\uparrow-z_j^\downarrow)^n,
\label{mmn}
\end{equation}
where $z_i^\uparrow$ and $z_i^\downarrow$ are the complex coordinates of
the $i$th electron in the upper and lower layer respectively. The common
exponential factors are neglected in (\ref{mmn}). The region with $V(r) < \mu$
is filled  
with the incompressible
electron liquid described by (\ref{mmn}), with Landau level filling factor
in individual layers $\nu_\uparrow=\nu_\downarrow={1\over m+n}$;
there is a gap for all bulk excitations which is of order
the smallest nonzero $U_l$\cite{gapnote}.
Gapless excitation can only live at
the edge of the incompressible liquid, which is along $r=R$ 
with $V(R)=\mu$ (see Fig. \ref{drop})\cite{edgenote}.

In such a disc geometry, there is equal amount of 
current flowing counter-clock-wisely 
along the edge in both layers, 
due to the gradient of the confining potential, while no 
current is flowing in the bulk. We may increase the edge current in the upper 
layer without changing the current in the lower layer by adding more charge to 
the upper layer, so that the edge in the upper layer moves from $R$ to
$R'>R$. In the region $R<r<R'$, there is no electron in the lower layer, and
the filling factor in the upper layer is $1/m$, which is {\em bigger} than that 
of the bulk, due to the absence of electron in the lower layer in the same 
region. 
The additional current is therefore 
\begin{equation}
\delta I_\uparrow={1\over m}{e\over h}[V(R')-V(R)],
\end{equation}
while the change of the chemical potential in the upper layer is clearly
\begin{equation}
\delta\mu_\uparrow=V(R')-V(R)=m{h\over e}\delta I_\uparrow.
\end{equation}
Even though the edge of the lower layer is still at $r=R$, the charge added 
in the upper layer also increases the chemical potential in the lower layer.
This is because if we were to 
add one more electron to the lower layer (at $r=R$), 
there
would be a charge with the amount $n/m$ moved from $r=R$ to $r=R'$ in the
upper layer, because
electrons in the lower layer are seen as nodal points by electrons 
in the upper layer, 
as described by the wave function
(\ref{mmn}), and electrons in the upper layer are 
pushed away from them. This means an
addition energy cost of ${n\over m}[V(R')-V(R)]$ for each electron added to
the lower layer, therefore
\begin{equation}
\delta\mu_\downarrow=n{h\over e}\delta I_\uparrow.
\end{equation} 
From the above we find the following linear-response equation:
\begin{equation}
\left(
\begin{array}{c}
\delta\mu_\uparrow\\
\delta\mu_\downarrow
\end{array}
\right) 
=
{h\over e}{\bf K}
\left(
\begin{array}{c}
\delta I_\uparrow\\
\delta I_\downarrow
\end{array}
\right)=
{h\over e}
\left(\begin{array}{cc}
m & n\\
n & m
\end{array}
\right)
\left(
\begin{array}{c}
\delta I_\uparrow\\
\delta I_\downarrow
\end{array}
\right),
\label{edgehall}
\end{equation}
which is equivalent to Eq. (\ref{result}).
The edge trans-Hall-resistance is nothing but the
off-diagonal matrix element of the above resistance matrix:
\begin{equation}
\rho^{xy}_{\uparrow\downarrow}=n{h\over e^2} > 0.
\end{equation}
Its positive sign is anomalous because it means if the chemical potential in 
the lower layer were held a constant, adding current in the upper layer induces
a change of current in the {\em opposite} direction (back flow) in the
lower layer; this is opposite to what normally happens in a drag experiment
with two electron layers\cite{gramila}. 

So far we have been considering the special case where the Landau level 
spacing is infinite and the Halperin wave functions describe the ground state
exactly. In the following we show that when these conditions are released, 
the results derived above are not altered for edge currents.

As long as the edge is reasonably sharp so that edge reconstruction does not
occur, the low energy physics is well described by the chiral Luttinger
Liquid theory\cite{wen2}. In this theory each edge component  
is described by a bosonic field $\phi_\sigma$.
In our case we have two components (upper or lower layer), and
$\sigma=\uparrow$ or $\downarrow$.
The edge electron density for each component is $\rho_\sigma(x)={1\over 2\pi}
\partial_x\phi_\sigma(x)$, and they satisfy the following 
commutation relation:\cite{wen2}
\begin{equation}
[\phi_\sigma(x), \rho_{\sigma'}(x')]
=i({\bf K}^{-1})_{\sigma\sigma'}\delta(x-x'),
\end{equation}
where ${\bf K}^{-1}$ is the inverse of the ${\bf K}$ matrix discussed above
(here we need to assume that ${\bf K}$ contains no zero eigen value and 
therefore its inverse is well defined).
The edge Hamiltonian is quadratic in $\rho$:
\begin{equation}
H={1\over 2}\sum_{\sigma\sigma'}\int{dx}V_{\sigma\sigma'}\rho_\sigma(x)
\rho_{\sigma'}(x),
\end{equation}
where $V_{\sigma\sigma'}$ is a nonuniversal, positive definite interaction
matrix that depends on the details of the edge confining potential as well as
electron-electron interactions at the edge.
From the continuity equation one obtains the edge current operator
\begin{eqnarray}
I_\sigma(x)&=&{e\over 2\pi}\dot{\phi}_\sigma(x)={e\over 2\pi}{1\over i\hbar}
[\phi_\sigma(x), H]\nonumber\\
&=&
{e\over h}\sum_{\alpha\beta}K^{-1}_{\sigma\alpha}V_{\alpha\beta}\rho_\beta(x).
\end{eqnarray}
and $\langle I_\sigma(x)\rangle=0$ in the ground state. Now we raise the edge 
electrostatic potential
in layer $\sigma$ by the amount $\delta v_\sigma$. This introduce the
following perturbation to the edge Hamiltonian: 
$\delta H=-e\sum_\sigma
\delta v_\sigma\int{dx}\rho_\sigma(x)$.
Solving the new Hamiltonian one obtains
$\langle\rho_\sigma(x)\rangle=e\sum_\beta V^{-1}_{\sigma\beta}\delta v_\beta$,
and 
\begin{equation}
\langle I_\sigma(x)\rangle={e^2\over h}\sum_{\sigma\beta} K^{-1}_{\sigma\beta}
\delta v_\beta.
\end{equation}
Interestingly, we find the result does {\em not} depend on $V$, which involves
microscopic details.
Reversing the matrix we obtain exactly Eq. (\ref{result}).
We therefore find that the edge trans-Hall-resistance depends on the
topological $K$ matrix of the bulk FQHE state {\em only}, 
and is therefore quantized.

So far we have been focusing on the drag effect of edge current. In reality,
current may flow both in the bulk and along the edge. Using continuity
condition, i.e., any gain (or loss) of current at the edge must be 
compensated by the current from the bulk, it is straightforward to show that
the drag coefficient must be the same for bulk and edge 
currents.\cite{notebe} Therefore our results should apply to driving 
current flowing both in the bulk and along the edges, and do not depend on the
details of the current distribution in the sample.

We have been focusing on gapped double layer FQHE in this paper. It
is plausible, however, that Hall drag should exist in compressible double
layer systems as well, if interlayer correlation exists.
In particular, by tuning some control parameters like the layer separation $d$,
it is possible to tune the system through a phase transition from a 
compressible state to a correlated double layer quantum Hall state.
Our results suggest that Hall drag is a useful way to probe such a phase
transition: As $d$ decreases from above the critical separation $d^*$, the
Hall drag resistivity should increase (due to
the interlayer correlation that
is building up), and reach the quantized value at $d=d^*$. Further 
decreasing $d$ should have no effect on the quantized value.

Very recently, {\em longitudinal} drag measurement has been performed on
a double layer system at filling factor $\nu_\uparrow
=\nu_\downarrow=1/2$ for each individual 
layers.\cite{lilly} In these systems the layer separation is larger although
fairly close to the critical layer separation $d^*$ below which the 
quantized double layer $(111)$ state forms, so the system is compressible.
It is found that the {\em longitudinal} drag resistance is much larger than 
that of zero magnetic field for the same system, although still much smaller
that the predicted {\em Hall} drag resistance of the (111) state 
at reasonable temperatures, and appears to stay finite in the zero temperature
limit. We note that since $d$ is larger but close to $d^*$, it is possible 
that the system has a sizable {\em Hall} drag resistance, which can be mixed
into the {\em longitudinal} drag measurement in a two-terminal setup. 
Since the Hall drag resistance can be so large (compared to 
longitudinal drag resistance), even a very small mixture can lead to 
a very big effect.
It is also observed\cite{lilly} that there is strong nonlinear current effect
at low temperatures in drag signal, possibly due to sample 
inhomogeneity.\cite{jpe} We note that since the system is close to the 
phase boundary at which the system becomes a bilayer quantum Hall state, it
is possible that the (111) state gets stabilized in certain regions of the
sample due to inhomogeneity. Since the quantum Hall state has very small 
longitudinal resistance, it is possible that most of the current flow through
these regions when the current is very low, leading to apparently large 
drag signal and nonlinear effects as the current increases (so that some of 
the current has to flow elsewhere).\cite{notesmg}

In summary, we have demonstrated the existence of Hall drag effect in 
correlated double layer FQHE systems, which may be used to detect interlayer
electron-electron correlation directly. The trans-Hall-resistivity tensor 
is shown to be quantized at zero temperature.
 
The author is indebted to Jim Eisenstein and Mike Lilly for stimulating his
interest in this problem and numerous informative discussions. 
He has also benefited greatly from discussions with Steve Girvin,
Jason Ho and Allan MacDonald. This work was supported by a Sherman Fairchild
fellowship.

\begin{figure}
\caption{Schematic illustration of a double layer FQHE liquid and its edge.
In the ground state the liquid is filled (in both layers) in the shaded region,
up to $r=R$. Chiral current flows around the edge counterclockwisely.
When more current is added in the upper layer, the edge in the upper layer 
moves outward to $R'$.
}
\label{drop}
\end{figure}


\begin{references}
\bibitem{gapnote} In the special case of $m=n$, there is a gap for
charged excitations but no gap for neutral excitation
[\onlinecite{wenzee,yang}]
in the 
bulk.
\bibitem{edgenote} For simplicity we consider disc geometry with a single
edge. In real systems there are usually two edges with electrons moving 
in opposite directions. It is trivial to generalize the analysis below to
that case, and the conclusion is the same.
\bibitem{notebe} In Ref. [\onlinecite{renn}], it is assumed that the 
current flows in the bulk. In reality, it appears that most of the current
actually flow along the edges.
\end{references}
\end{document}